\documentclass{aa}
\usepackage[dvips]{graphicx}
\usepackage{txfonts}
\begin{document}
\title {The role of magnetic reconnection on jet/accretion disk systems}
\author {E.~M.~de Gouveia Dal Pino
				\inst{1}
				,
				P.~P.~Piovezan
				\inst{1, 2}
				\and
				L.~H.~S.~Kadowaki
				\inst{1}
				}
\institute{Universidade de S\~ao Paulo, IAG, Rua do Mat\~ao 1226, Cidade Universit\'aria, S\~ao Paulo 05508-900, Brazil\\
				\email{dalpino@astro.iag.usp.br}
\and
					Max Planck Institute for Astrophysics, Karl-Schwarzschild-Str. 1, Postfach 1317, D-85741 Garching, Germany
					}
\date{Accepted ??? ???. Received ??? ???; in original form ??? ???}
\abstract
{It was proposed earlier that the  relativistic ejections observed in microquasars could be produced by violent magnetic reconnection episodes at the inner disk coronal region (de Gouveia Dal Pino \& Lazarian 2005). }
{Here we revisit this model, which employs a standard accretion disk description and fast magnetic reconnection theory, and discuss the role of magnetic reconnection and associated heating and particle acceleration in different jet/disk accretion systems, namely young stellar objects (YSOs), microquasars, and active galactic nuclei (AGNs).}
{In microquasars and AGNs, violent reconnection episodes between the magnetic field lines of the inner disk region and those that are anchored in the black hole are able to heat the coronal/disk gas and accelerate the plasma to relativistic velocities through a diffusive first-order Fermi-like process within the reconnection site that will produce intermittent relativistic ejections or plasmons.}
{The resulting power-law electron distribution is
compatible with the synchrotron radio spectrum observed during the outbursts of these
sources. A diagram of the magnetic energy rate released by violent reconnection as a
function of the black hole (BH) mass spanning $10^9$ orders of magnitude shows that
the magnetic reconnection power is more than sufficient to explain the observed radio
luminosities of the outbursts from microquasars to low luminous AGNs. In addition,
the magnetic reconnection events cause the heating of the coronal gas, which can be
conducted back to the disk to enhance its thermal soft x-ray emission as observed
during outbursts in microquasars.  The decay of the hard x-ray emission right after a radio
flare could also be explained in this model due to the escape of relativistic electrons
with the evolving jet outburst. In the case of YSOs a similar magnetic configuration
can be reached that could possibly produce observed x-ray flares in some sources and provide
the heating at the jet launching base, but only if violent magnetic reconnection events occur
with episodic, very short-duration accretion rates which are $\sim 100-1000 $ times
larger than the typical average accretion rates expected for more evolved (TTauri) YSOs.
%This study is  a first step and  multi-dimensional
%numerical models will be necessary to confirm the present results.
}
{}
\keywords{Magnetic reconnection -- Accretion disks -- Relativistic Jets -- Microquasars and AGNs -- YSOs}
\authorrunning{E.~M.~de Gouveia Dal Pino et al.}
\maketitle
\section{Introduction}
Supersonic jets are observed in several astrophysical systems, such as
active galactic nuclei (AGNs), neutron star and black hole X-ray binaries, and low-mass young stellar objects (YSOs), and are probably also associated with gamma-ray bursts.  The study of their origin, structure, and evolution
helps to shed light on the nature of their compact progenitors, as they
carry angular momentum, mass, energy, and magnetic field away from
the sources.

The currently most accepted paradigm for jet production is based on the
magneto-centrifugal acceleration out of a magnetized accretion disk
that surrounds the central source. Firstly proposed by  Blandford \& Payne (1982; see also Lynden-Bell 1969; Blandford \& Rees 1974; and Lovelace
1976, where these ideas initially germinated), this basic scenario for jet launching has been object of extensive analytical and numerical investigation (see e.g.,  McKinney \& Blandford 2009; Shibata 2005; de Gouveia Dal Pino 2005 for reviews). However, though considerable progress has been achieved in the comprehension of the possible origin of the
magnetic fields that must permeate the accretion disk and the mechanism
of angular momentum transport that allows the accretion to occur through magnetorotational turbulence (Balbus \& Hawley 1998), questions related to jet stability, the nature of the coupling between the central source magnetosphere
and the disk field lines, and the quasi-periodic ejections that are
often associated to these jets, are  still debated.

Relativistic jets from stellar-mass black holes of binary stars
emitting X-rays, also called microquasar jets (or
BHXRB jets), are scaled-down versions of AGN (or quasar) jets,
typically extending for $\sim$ 1pc and probably powered by the accreting,
spinning black hole. Despite the enormous difference in scale, both
classes share several similarities in their physical properties.
However, because the characteristic times of the matter flow are
proportional to the black hole mass, the accretion-ejection
phenomena in microquasars end sooner and are about $10^{-7} - 10^{-5}$
faster than analogous phenomena in quasars. This fact and their
closer proximity to us (they are generally observable within the
Galaxy) make the microquasars easier to investigate compared to the
AGNs (Mirabel \& Rodrigues 1999).

Although individual systems are complex and peculiar when looked at in detail, there are common features to all classes of BHXRBs (e.g, Remillard \& McClintock, 2006). According to their x-ray emission (2-20 keV), they show basically two major states: a
quiescent and an outburst state. The former is characterized by low
x-ray luminosities and hard non-thermal spectra. Usually, transient
BHXRBs exhibit this state for long periods, which allows one to obtain
the physical parameters of the system as the spectrum of the
secondary star becomes prominent.

On the other hand, the outburst state corresponds to intense
activity and emission and can be sub-classified in three main active
and many intermediary states. According to Remillard \& McClintock
(2006), the three main active states are the thermal state (TS), the
hard state (HS) and the steep power law state (SPLS). These states
are usually explained as changes in the structure of the accretion
flow. During the TS, for example, the soft x-ray thermal emission
comes from the inner region of the thin accretion disk that extends
until the last stable orbits around the black hole. On the other
hand, during the HS the observed weak thermal component suggests
that the disk is truncated at a few hundreds/thousands gravitational
radii. The hard x-ray emission measured during this state is often
attributed to inverse Compton scattering of soft photons from the
outer disk by relativistic electrons in the hot inner region of the
system (e.g. Remillard \& McClintock, 2006; Malzac, 2007).

A widely observed example from radio to x-rays is the microquasar
GRS 1915+105. Located at a distance of $\sim 12.5$ kpc and probably
with a 10--18 solar mass black hole in the center of the binary
system, it was the first galactic object to show evidence of
superluminal radio ejection (Mirabel \& Rodrigues, 1994; Mirabel \&
Rodrigues, 1998). Dhawan et al. (2000) have distinguished two main
radio states of this system, a plateau and a flare state. During the
plateau state the RXTE (2--12 keV) soft X-ray emission is weak,
while the BATSE (20--100 keV) hard X-rays are strong and the radio
flat spectrum is produced by a small scale nuclear jet. On the other
hand optically thin ejecta are superluminally expelled up to thousands of AU during the flare phase and the radio spectral
index is between 0.5 and 0.8. The soft X-rays also flare during this
phase and exhibit a high variability, while the hard X-rays fade for a
few days before recovering. In terms of the x-ray spectral states,
several works have verified that the radio flares of this source
occur during the SPLS (e.g., Fender et al., 2004) and the x-ray
emission of both the plateau and the flare state are different
manifestations of the SPLS (e.g., Reig et al., 2003).

The examination  of  radio and x-ray observations for other
microquasars (e.g., XTE J1859+226, XTE J1550-564) also suggests that
the ejection of relativistic matter happens when the source is very
active and in the SPLS (e.g., Hannikeainen et. al, 2001; Brocksopp
et. al, 2002; Fender et al., 2004; McClintock et. al, 2007).
According to Fender et al. (2004), the origin of the optically thin
emission in radio is due to shock waves that are formed in the jet
when the system passes from a `hard' SPLS to a `soft' SPLS. However,
what generates these shock waves or the triggering mechanism for the
ejections of matter is unclear.

In 2005, de Gouveia Dal Pino \& Lazarian (see also de Gouveia Dal Pino 2006) proposed a mechanism that
can be responsible for the initial acceleration of the plasma jet to
relativistic speeds in the case of  the microquasar GRS1915+105.
Their scenario is related to violent reconnection episodes between
the magnetic field lines of the inner disk region and those that are
anchored in the black hole. We here revisit this model
and argue that it could be responsible for the transition from the
`hard' SPLS to the `soft' SPLS seen in other microquasars. We also
extend it to the AGN jets and briefly discuss its role for thermal YSO jets (see also de Gouveia Dal Pino 2006 and de Gouveia Dal Pino et al. 2009 for earlier discussions).

 We note that an interesting correlation between the radio and the (hard) X-ray luminosity has been reported by Laor \& Behar (2008) for sources spanning $10^{10}$ orders of magnitude in mass (from magnetically active stars to some galactic black holes and radio quiet AGNs), and recently extended to dwarf nova outbursts by Soker \& Vrtilek (2009). This correlation holds only for sources for which
 $L_R/L_X < 10^{-5}$
and suggests that a common mechanism may be responsible for  the radio emission in these highly different objects. Because in magnetically active stars the emission is related to the coronae, these authors have argued that the same might be occurring in some BHXRBs and radio quiet AGNs, i.e.,  the radio emission could also come from magnetic activity in the coronae above the accretion disk, just like in the scenario that we investigate here in detail (and which was previously suggested in de Gouveia Dal Pino 2006, and de Gouveia Dal Pino et al. 2010a, 2010b).

Earlier studies have also revealed a correlation between the radio and the X-ray luminosity of BHXRBs in their quiescent "low/hard" state and low-luminosity AGNs
 with nuclear radio emission or weak radio jets  (Gallo et al. 2003;
Merloni et al. 2003; Falcke et al. 2004).
\footnote{We notice that
these correlations are often interpreted in the context of
(advection dominated accretion flow) ADAF models  where
a  geometrically thick and optically thin accretion disk  with small mass flow rate
is invoked to explain the low/hard X-ray emission accompanied by radio jet activity.
 In these models a geometrically thin disk structure in the inner region would be recovered when the system switches to a high-accretion flow rate, hence producing  a high/soft emission with suppression of
the jet activity. In the present model we examine these objects mostly during their
outburst phases rather than during their more quiescent states. Besides, because the reconnection mechanism
is operating at the interface region between the inner disk and the central source, the
choice of a geometrically
thin accretion disk (rather than a  thick one) is not a crucial point for our model.}

As also remarked by Soker \& Vrtilek, suggestions for the presence of coronae above accretion disks are not new (e.g., Liang \& Price 1977, Liang 1979, Galeev et al. 1979; Done \& Osborne 1997; Wheatley \& Mauche 2005; Cao 2009), and the connection between coronae and jets were previously proposed as well (e.g., Fender et al. 1999; Markoff et al. 2005; de Gouveia Dal Pino \& Lazarian 2000, 2001, 2005).
If the magnetic activity in erupting accreting disks is similar to that in stars, then coronal
mass ejections,  as in the Sun, could be expected. Hence magnetic
flares similar to those in active stars might be a potential mechanism operating at the jet launching region in a variety of systems, from young stellar objects to BHs (de Gouveia Dal Pino 2006, de Gouveia Dal Pino et al. 2009; Soker 2007;  Soker \& Vrtilek 2009).

%The
%idea that jet outbursts may be launched by magnetic fields reconnection
% events similar to those in the sun is not new, e.g., de Gouveia Dal Pino \&
%Lazarian (2005; de Gouveia Dal Pino 2006; de Gouveia Dal Pino et al. 2008).

This paper is organized as follows: in Sect. 2 we briefly describe
de Gouveia Dal Pino \& Lazarian's model. In Sect. 3 we discuss it
in the context of the microquasars and in Sect. 4 we propose its generalization to
other classes of relativistic jets, extending it to low-luminosity
AGN jets. In Sect. 5 we briefly discuss the reconnection model in
the context of the YSO jets and finally, in Sect. 6, we draw our
conclusions.

\section{A simple reconnection scenario in the inner accretion disk/corona region}
We consider a magnetized accretion disk around a rotating (Kerr) BH as schematized in Fig. 1. A detailed description of
the scenario adopted is given in de Gouveia Dal Pino \& Lazarian
(2005). Here we briefly present the assumptions  made.

\begin{figure}
 \begin{center}
  \includegraphics[width=0.45\textwidth]{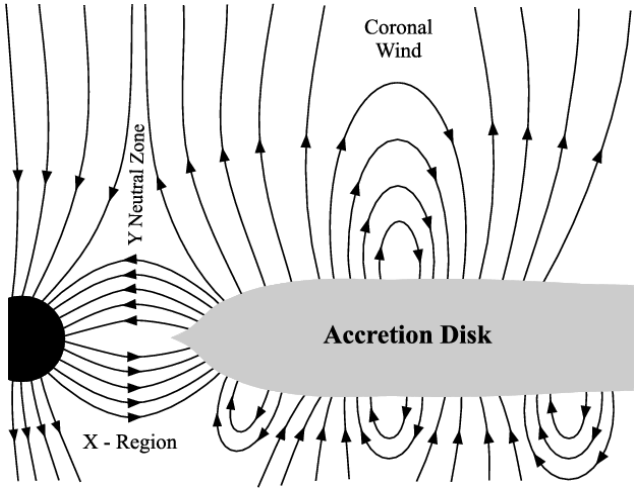}
  \caption{Schematic drawing of the magnetic field geometry in the inner disk-source  region at $R_X$. From de Gouveia Dal Pino \& Lazarian (2005).}
  \label{fig:fig1}
 \end{center}
\end{figure}

A magnetosphere around the central BH may be formed from the drag of
magnetic field lines by the accretion disk (e.g., MacDonald et al. 1986; Wang et al. 2002).
The disk large-scale poloidal magnetic field could be established
%\textbf{either by advection of magnetic field from the outer to the inner parts of the accretion disk (e.g., Beckwith, Hawley, Krolik 2009) or}
by the action of a turbulent dynamo inside the accretion disk
(Livio et al. 2003; King et al. 2004; Uzdensky \& Goodman 2008) driven, e.g., by  the magnetorotational
instability (Balbus \& Hawley 1998) and disk differential rotation.
Although this scenario remains speculative and nothing definitive has been revealed yet by  global disk numerical simulations,
%We further assume that once
%the dynamo process establishes a global poloidal field over a
%substantial region of the disk, it will be able to maintain that
%field for a period of time large enough for the process we will
%describe below to occur.
the action of a turbulent dynamo followed by the advection of the
magnetic flux with the gas towards the inner disk region could result
in a gradual increase of the magnetic flux in this region (e.g.,
Merloni 2003; Livio et al. 2003; Tagger et al. 2004; de Gouveia Dal
Pino \& Lazarian 2005; McKinney \& Blandford 2009).
Even without any dynamo action, a net poloidal field could be built in the inner disk-corona region simply due to advection of existing magnetic flux from the large radii. This has been recently proven to be plausible through three-dimensional numerical simulations (see, e.g., Beckwith, Hawley and Krolik 2009 and references therein). These authors have found in particular that most of the magnetic flux inward motion should operate outside the disk body in the coronal region. They found  that  magnetic stresses driven by differential rotation can create large-scale half-loops of magnetic field  that stretch
radially inward and then reconnect, leading to discontinuous jumps in the location
of magnetic flux. At the end, this mechanism establishes a dipole magnetic field in the inner region with intensity  regulated by a combination of  magnetic and gas pressures in the inner disk. Although these numerical studies are still preliminary and the saturation condition for this mechanism is not fully understood yet, these results are a demonstration that it is possible to establish a large-scale poloidal field in the inner disk-coronal region.

According to the magneto-centrifugal scenario (Blandford \& Payne 1982), this
poloidal magnetic flux summed to the disk differential rotation can
give rise to a wind that removes angular momentum from the system
and increase the accretion rate. This will increase the ram pressure
of the accreting material that will then push the magnetic lines in
the inner disk region towards the lines which are anchored in the BH
horizon allowing a magnetic reconnection event to occur (see the Y-type zone labeled
as helmet streamer in Fig. 1). Also, with the accumulation of the
poloidal flux in the inner regions the ratio between the
gas$+$radiation pressure to the magnetic field pressure ($\beta$)
will soon decrease. \footnote{We will see below that $\beta$ is left as a free parameter in our model, but for violent reconnection events to be produced it must actually decrease to values lower than unity.} As we show in Sect. 3.2, when the accretion
rate reaches values close to the Eddington limit and $\beta < 1$,
the magnetic reconnection event becomes violent and a substantial amount of magnetic energy
is released quickly by this process.

If the BH is rotating, closed field lines connecting
it with the inner disk edge can lead to  angular
momentum transfer between them if they
are rotating at different angular speeds
(Blandford-Znajek 1977). To facilitate the analysis, we consider as in de Gouveia Dal Pino \& Lazarian, that the BH and the inner zone of the accretion disk are nearly co-rotating in a way that there is no net angular momentum and energy transfer between them. This is strictly speaking not a necessary condition of the model.

\section{Revisiting the model: microquasar ejections}
We assume here that during the plateau state that precedes a radio
flare (e.g., in the microquasar GRS 1915+105) the large scale
poloidal field is progressively built in the disk by the dynamo
process. Immediately before the flare, the idealized configuration of the
system is that of Fig. 1 and  both the accretion rate and the
magnetic pressure in the inner disk zone become very high  due to
the reasons described in the preceding section. In order to evaluate the
amount of magnetic energy that can be extracted through violent
reconnection, we first need to evaluate the physical conditions both
in the magnetized disk and in the corona that surrounds it.

\subsection{Disk and coronal parameters}
Let us assume as in de Gouveia Dal Pino \& Lazarian (2005) that the inner radius of the accretion disk ($R_X$)
approximately corresponds to the last stable orbit around the BH.
For a BH with stellar mass $M \simeq  14 M_{\odot} = M_{14}$ and
Schwartzschild radius $R_S = 2GM/c^2 = 4.14 \times 10^6 M_{14}$ cm
(which are parameters suitable for instance for GRS1915+105), this gives
$R_X = 3 R_S \sim 10^7$ cm.

X-rays observations show that the internal regions of microquasar
accretion disks have temperatures higher than $\sim 10^6$ K. At
these temperatures, hydrogen, which is probably the main constituent of
the disk material, is completely ionized. In this case the main
source of disk opacity is the Thomson scattering. Moreover, the
thermal pressure of the gas is much less than the radiation
pressure and can be neglected. To evaluate the disk quantities at
$R_X$, we adopt the standard  model (Shakura \& Sunyaev 1973).

 In this case, the set of equations that describe a geometrically thin,
optically thick, stationary accretion-disk with a Keplerian profile
 is given by
 \footnote{We notice that the standard Shakura-Sunyaev model with an $\alpha$ prescription for the viscous stress (which is assumed to be proportional to the total pressure, with $\alpha$ the constant of proportionality) is  an idealized simplification of the real physics in accretion disks.  When in the radiation dominated regime, it is well known to be subject to thermal and inflow (viscous) instabilities, which make a stationary solution unrealistic. Nonetheless, Hirose and collaborators (see Hirose, Krolik, \& Stone
2006; Hirose, Krolik \& Blaes 2009; Hirose, Blaes \& Krolik 2009, and references therein), for instance,  have recently explored numerically more realistic models, and although they found inconsistencies in the $\alpha$-models, like for instance that these seem to be actually thermally stable, they could not  provide better analytical solutions. We further remark that the real structure of the accretion disk is not a crucial
point for the purposes of the present study because the focus is the inner disk-coronal region where the interaction with the central source takes place.}

\begin{equation}
T_{d} \cong 1,64 \times 10^{7} \alpha_{0.5}^{-1/4} M_{14}^{1/8}
R_{X,7}^{-3/8} \hspace{0.2cm} K
\end{equation}
\begin{equation}
n_{d} \cong 3,65 \times 10^{18} \alpha _{0.5}^{-1} M_{14}^{-1/2}
\dot{M}_{19}^{-2} R_{X,7}^{3/2} q_{0.82}^{-8} \hspace{0.2cm} cm^{-3}
\end{equation}
\begin{equation}
H_{d}/R_{X} \cong 0,57 \dot{M}_{19} R_{X,7}^{-1} q_{0.82}^{4}
\end{equation}
\begin{equation}
U_{d} \cong 4,91 \times 10^{14} \alpha _{0.5}^{-1} M_{14}^{1/2}
R_{X,7}^{-3/2} \hspace{0.2cm} erg/cm^{3},
\end{equation}
where $\alpha = 0.5\alpha_{0.5}$ is the disk viscosity, $M = 14
M_{\odot} M_{14}$, $R_{X} = 10^{7} R_{X,7}$, $\dot{M} = 10^{19}
\dot{M}_{19}$ is the disk mass accretion, $q =
[1-(R_{S}/R_{X})^{1/2}]^{1/4} =0.82 q_{0.82}$, $T_d$ is the disk
temperature, $n_d$ its density, $H_d$  its half-height, and
$U_d$ is the photon energy density emitted by the disk.

To determine the accretion rate immediately before an event of
violent magnetic reconnection, we assume  the equilibrium between
the disk gas ram pressure and the magnetic pressure of the magnetosphere
anchored at the event horizon of the black hole.
\footnote{A note is in order here. As remarked in Sect. 2, we assume according to the magneto-centrifugal scenario that the
poloidal magnetic flux established in the inner disk region summed to the disk differential rotation can
give rise to a wind that removes angular momentum from the system, thus increasing the accretion rate. This will increase the ram pressure of the accreting material (which is continuing to flow from the outer parts of the disk), which will then further push the magnetic lines in the inner disk region towards the lines which are anchored in the BH horizon, allowing a violent reconnection event. Based on this, we can assume the balance between the magnetic pressure and the disk gas ram pressure right before an event of violent reconnection in the inner disk region. On the other hand, without a continuous inflow from the outer disk regions, one might guess that the wind could greatly reduce the amount of accretion flow that reaches the inner disk radius.

 A similar effect could be achieved also in an ADIOS (adiabatic dominated inflow outflow)  model (Blandford \& Begelman 1999), which is an ADAF-like solution (Narayan \& Yi 1994) with an outflow. However, these  are appropriate solutions only for systems that accrete at rates well below the Eddington rate, which is not the case discussed here.
Of course, in a real system any of the possibilities above could occur, but in the present study we seek for the best conditions under which a violent reconnection event can be achieved, using a simple analytical description  to obtain upper limit estimates for the amount of energy that can be extracted from the reconnection event. Given the non-linearity and intrinsic non-steadiness character of the problem, we hope that future numerical simulations  will be able to give us more realistic evaluations.}

 Assuming spherical geometry, the radial accretion velocity can be approached by the
free fall velocity. Also, assuming that the intensity of the field
anchored in the BH horizon is on the order of the inner disk
magnetic field (MacDonald et al. 1986; de Gouveia Dal Pino \&
Lazarian 2005), we find that
\begin{equation} \label{eq:eqRamMag}
\frac{3 \dot{M}}{4 \pi R^2}  \left( \frac{2GM}{R} \right)^{1/2} \sim \frac{B_{d}^2}{8 \pi}.
\end{equation}

On the other hand, the disk magnetic field can be parameterized by
$\beta$, here defined as the ratio between the  gas+radiation pressure (which is
dominated by the radiation pressure) and the magnetic pressure. Using Eq.(4) to obtain the radiation pressure ($p_{d} = U_{d}/3$) and the equation above,  we find that the magnetic field and the accretion
rate at $R_X$ are
\begin{equation} \label{eq:Bd}
B_d \cong 7.54 \times 10^{7} \beta_{0.8}^{-1/2} \alpha _{0.5}^{-1/2} M_{14}^{1/4} R_{X,7}^{-3/4} \hspace{0.2cm} G,
\end{equation}
\begin{equation} \label{eq:Rate}
\dot{M} \sim 1 \times 10^{19} \beta_{0.8}^{-1} \alpha_{0.5}^{-1} R_{X,7} \hspace{0.2cm} g/s,
\end{equation}
where $\beta = 0.8 \beta_{0.8}$.

Liu et al. (2002) proposed a simple model to quantify the parameters
of the corona of a magnetized disk system. Assuming as in the solar
corona that gas evaporation at the foot point of a magnetic flux
tube quickly builds up the density of the corona to a certain value
and that the tube radiates the heating due to magnetic reconnection
through Compton scattering, the coronal temperature and density in
terms of the disk parameters can be derived respectively as
\begin{equation} \label{eq:Tc}
T_{c} \cong 4,10 \times 10^{9} \alpha_{0.5}^{-1/8} \beta_{0.8}^{-3/8} M_{14}^{1/16} R_{X,7}^{-3/16} l_{100}^{1/8} \hspace{0.2cm} K
\end{equation}
\begin{equation} \label{eq:nc}
n_{c} \cong 5,36 \times 10^{15} \alpha_{0.5}^{-1/4} \beta_{0.8}^{-3/4} M_{14}^{1/8} R_{X,7}^{-3/8} l_{100}^{-3/4} \hspace{0.2cm} cm^{-3},
\end{equation}
where $l_{100}= 100 R_X$ is the scale height of the Y neutral
zone in the corona. In the equations above we also assumed that
the magnetic field at the reconnection region in the corona is on the order of the
field anchored at the disk inner radius (see Fig. 1).

% Inserir nas eqs. acima a dependencia de B com a distancia
%l^-\eta (com o indice da potencia variando de 0 a 2; 0 significa que
%B = B_d.

\subsection{Rate of magnetic energy released by magnetic reconnection}
As described in de Gouveia Dal Pino \& Lazarian (2005), the rate of
magnetic energy that can be extracted from the Y-zone in the corona
(above and below the disk) through reconnection is
\begin{equation} \label{eq:WB}
\dot{W}_{B} \approx \frac{B^{2}}{8 \pi} v_{A} (4 \pi R_{X} \Delta R_{X}),
\end{equation}
where $B$ is the magnetic field at the reconnection zone, $v_A$ is
the coronal Alfv\'en speed, $v_A = B/(4\pi n_c m_p)^{1/2}$, $m_p$
the hydrogen mass, and $\Delta R_{X}$ is the width of the current
sheet. This last term can be estimated considering the condition for
which the resistivity at the reconnection zone is anomalous, as  in
Lazarian \& Vishiniac (1999) and de Gouveia Dal Pino \& Lazarian
(2005):
\begin{equation}\label{eq:DR}
%\begin{eqnarray} 
\left(\frac{\Delta R_{X}}{R_{X}}\right) \cong 1,86 \times 10^{-5}
Z^{-1} \alpha_{0.5}^{-3/16} \beta_{0.8}^{7/16} M_{14}^{3/32}
%\nonumber \\ && \times
R_{X,7}^{-41/32} l_{100}^{11/16}
\end{equation}
%\end{eqnarray}

 From Eqs. \ref{eq:Bd}, \ref{eq:WB}, and \ref{eq:DR} we
estimate the amount of magnetic energy that can be extracted as
\begin{equation} \label{eq:MicroPot}
\dot{W}_{B} \cong 1,6 \times 10^{35} \alpha_{0.5}^{-19/16} \beta_{0.8}^{-9/16} M_{14}^{19/32} R_{X,7}^{-25/32} l_{100}^{11/16} \hspace{0.2cm} erg/s.
\end{equation}

 The corresponding reconnection time is
\begin{equation} \label{eq:Microtempo}
t_{rec} \cong \frac{R_{X}}{\xi v_{A}} \cong 10^{-4} \xi^{-1} R_{X,7} \hspace{0.2cm} s,
\end{equation}
where $\xi = v_{rec}/v_A$  is the reconnection rate and $v_{rec}$ is
the reconnection velocity. For the conditions investigated here we
find that $v_A \simeq c$.

This relation indicates that for efficient reconnection the release
of magnetic energy is very fast, as is that required to produce flares.
\footnote{We note that solar flare observations indicate that the
reconnection speed can be as high as a few tenths of the Alfv\'en
velocity (e.g., Takasaki et al. 2004).} The energy released could be
used to heat the coronal gas (as required by the coronal model; see
Eq. \ref{eq:nc}) and also to accelerate particles to
relativistic velocities, producing a violent radio ejecta (as we will
see in the next section).

After reconnection, the destruction of the vertical magnetic flux in
the inner disk will increase $\beta$, and  the corona will return to
a less magnetized condition with most of the energy dissipated
locally in the disk, instead of in the outflow.

\subsection{Particle acceleration and radio emission}
According to the discussion in the previous section, we argue that
whenever $\beta$ (the ratio between gas+radiation pressure and
magnetic pressure) becomes small enough and the accretion rate
attains values near the critical one in the inner disk region, the
lines of opposite polarization near the Y-zone will be pressed
together by the accreted material sufficiently rapidly as to undergo
violent reconnection events. Then energy stored in the magnetic
field will be released suddenly and at least part of it will be used
to accelerate charged particles. High-speed particles will spew outward, cause the relativistic radio ejecta and produce a luminous blob.

de Gouveia Dal Pino \& Lazarian (2005) have proposed a mechanism to
accelerate particles to relativistic velocities within the
reconnection zone in a  similar process to the first-order Fermi.
Charged particles are confined in the reconnection zone by the
particle+magnetic flux coming from both sides of the current sheet
in a way that their energies increase stochastically. They  have shown that a power-law
electron distribution with $N(E) \propto E^{-5/2}$ and a synchrotron
radio power-law spectrum $S_{\nu} \propto \nu^{-0.75}$ can be
produced in this case. Kowal, de Gouveia Dal Pino \& Lazarian are
presently testing this acceleration model in reconnection sites
using 3D Godunov-MHD simulations combined with a particle in-cell
technique (Kowal et al., 2009, in prep.).

This mechanism does not remove the possibility that further out the
relativistic fluid may also be produced behind shocks, which are
formed by the magnetic plasmons that erupt from the reconnection
zone. Behind these shocks a standard first-order Fermi acceleration
may also occur resulting a particle power-law spectrum $N(E) \propto
E^{-2}$ and a synchrotron spectrum $S_{\nu} \propto \nu^{-0.5}$.
Both radio spectral indices are consistent with the observed
spectral range during the flares of GRS 1915-105 ($-0.2 < \alpha_{R} <
-1.0$; Dhawan et al., 2000; Hannikainen et al., 2001).

\subsection{X-ray emission}
The enhanced x-ray emission that often accompanies the violent
flares in microquasars can be easily explained within the scenario
presented here as due to the increase in the accretion rate
immediately before the violent magnetic reconnection events.

The observed soft x-ray emission is expected to be a fraction of the
accretion power, and for an enhanced accretion rate near the
Eddington rate this is given by
\begin{equation} \label{eq: MicroWac}
\dot{W}_{ac} \cong \frac{G M_{BN} \dot{M}}{R_{X}} \cong 1.87 \times 10^{39} M_{14} \dot{M}_{19} R_{X,7}^{-1} \hspace{0.2cm} erg/s ,
\end{equation}
which is compatible with the soft x-rays observations of Remillard \& McClintok (2006). On
the other hand, the hard x-ray component, which is also often observed,
can be explained by inverse Compton scattering of the soft x-rays
photons by the hot electrons of the corona/jet. If this is the case,
we expect that after the radio flare the hard x-ray luminosity
will decrease because of the decrease of the number of
relativistic electrons (because most of them are accelerated away from
the system). Indeed, this behavior is also compatible with the
observations (e.g. Dhawan et al, 2000).

An  estimation of the electrum spectrum at the launching region produced in a magnetic reconnection  episode indicates that most of the relativistic electrons should be self-absorbed at this point.   When moving away, the produced plasmon cloud dilutes,
becoming transparent to its own radiation, first in the infrared and
then in radio frequency. The computation of the evolution of this
spectrum (e.g., Reynoso \& Romero 2009) is out of the scope of the present work as it requires the
building of a detailed model for the corona, which will be considered
elsewhere.
Nonetheless, we can make some predictions by comparing
the calculated power released during a violent magnetic reconnection event with the
observed luminosities, as below.

\subsection{Parameter space}

Figure 2 shows a comparison between the calculated magnetic power released during a
violent magnetic reconnection event (Eq. 12) and the observed luminosities for three microquasars with radio jet production, namely GRS1915+105 (a), XTE
J1859+226 (b), and XTE J1550-564 (c).
% The stars and circles represent their observed radio and infra-red luminosities, respectively.
The diagram shows the calculated magnetic power
released in violent magnetic reconnection events as a function of the central source
mass. A suitable choice of the parameter space was used in this
calculation: $5 M_{\odot} \leq M \leq 20 M_{\odot}$ (e.g., Remillard
\& McClintock, 2006), $0.05 \leq \alpha \leq 0.5$ (e.g., King et al.,
2007), $0.1 \leq \beta \leq 1$, $1 R_S \leq l \leq 1000 R_S$.

\begin{figure}
 \begin{center}
 \includegraphics[width=0.47\textwidth]{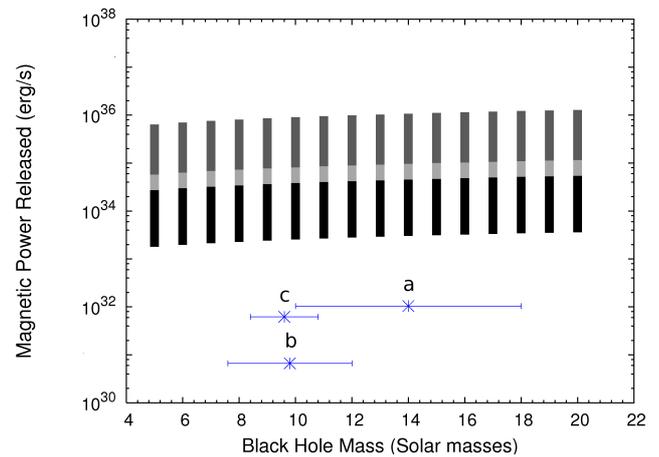}
 \caption{$\dot{W}_B$ as function of the microquasar black hole mass $M$. The (blue) stars  represent the observed  radio luminosities for three microquasars: GRS 1915+105 (a), XTE J1859+226 (b), and XTE J1550-564 (c). The gray scale bars correspond to the calculated magnetic reconnection power and encompasses the parameter space that spans $5 M_{\odot} \leq M \leq 20 M_{\odot}$, $0.05 \leq \alpha \leq 0.5$, $0.1 \leq \beta \leq 1$, and
 $1 R_S \leq l \leq 1000 R_S$ (or $0.3 R_X \leq l \leq 333 R_X$), with $1 R_S \leq l \leq 10 R_S$ in black; $10 R_S < l \leq 30 R_S$ in light gray; and $30 R_S < l \leq 1000 R_S$ in dark gray.}
 \end{center}
\end{figure}

Figure 2 indicates that the magnetic power released by violent
magnetic reconnection events is high enough to explain the radio
flare luminosities and just marginally high enough to account for the
IR luminosity (which has values between $10^{34}$ erg $s^{-1}$ and $10^{36}$ erg $s^{-1}$,
for the sources  XTE J1859+226 (b) and GRS1915+105 (a), respectively).
%In fact, if the coronal electrons are accelerated to
%relativistic speeds at the reconnection zone then one expect that
%the synchrotron radio luminosity produced by these electrons would
%be smaller than the power $\dot{W}_B$ available to accelerate them.
It is also possible that additional  emission results from
electrons that were re-accelerated in optically thin regions away from the
reconnection zone (behind shock waves along the jet excited by the
plasmon clouds, as emphasized above). This idea is supported for instance by IR
observations of GRS 1915+105, which indicate a separation between
the IR emission and the central source $\sim 0.2$ arcsec or $\sim
2500$ AU at 12.5 kpc (Sams et al. 1996), ie., further away from the predicted launching region by the present model (Fig. 2).

Finally,
%within the
%scenario presented here the soft x-ray emission can be larger than
%$\dot{W}_B$ once its origin is not directly related to the magnetic
%reconnection process (as discussed in the previous section). Also,
because the magnetic power released by reconnection is much
greater than the observed radio luminosity, only a
small fraction of this energy is necessary to accelerate particles. Then
most of the energy released would be used to heat the coronal and
also the disk gas surface (by thermal conduction along the field
lines). This supplementary heating could perturb the accretion disk and result in the
variability of the soft x-ray emission observed during the flare
phase of GRS 1915+105 (Dahwan et al., 2000).

\subsection{Transition between states}
According to the scenario presented here, Fig. 1 would
correspond to the configuration of the system immediately before a
radio flare. Thus, this could be the system configuration at the end
of the 'hard' SPLS and the transition from  the 'hard'  to the
'soft' SPLS would be due to violent magnetic reconnection events
between the magnetic field lines of the inner disk region and those
that are anchored into the black hole.
%These reconnection events
%release great amounts of energy (whenever $\dot{M} \lesssim
%\dot{M}_{Edd}$ and $\beta \lesssim 1$) that could be used to
%accelerate coronal particles producing the relativistic ejections of
%matter.

It is important to emphasize that the physical origin of the hard
x-ray emission in the SPLS is yet controversial in the literature.
Most models claim for the inverse Compton scattering as the dominant
radiative mechanism (e.g. Zdziarski, 2000), and photons with MeV
energies suggest that the scattering occurs in a non-thermal corona
(e.g. Zdziarski, 2001). For Poutanen \& Fabian (1999), the origin of
scattering electrons could be the coronal active magnetic regions
that rise from the disk, which agrees with the scenario proposed
here. Furthermore, the SPLS tends to dominate the spectrum of the
BHXRBs as the luminosity reaches values near the Eddington rate, which
also agrees with the model presented here.

\section{Generalizing the model: AGNs ejections}
Despite the huge difference in scales, AGNs/quasars and microquasars
have similar morphologies (Mirabel \& Rodrigues 1998). Indeed, some studies
indicate that the similarity between these systems is more than
morphological: they are possibly subject to the same physical
processes (e.g., de Gouveia Dal Pino 2005 and references therein for a review). If this is the case, then a generalization of the above
scenario to the extragalactic sources is almost straightforward. Indeed, relativistic jets from AGNs are also observed to produce
relativistic episodic ejections that originate synchrotron power-law
spectrum with similar spectral indices as above.

Considering the same assumptions presented above, we obtain similar
scaling relations for AGNs. Figure 3 depicts a synthesis of the
generalization of the magnetic reconnection scenario for
relativistic sources including both microquasars and AGNs. The
diagram shows the calculated magnetic power released in violent
magnetic reconnection events as a function of the central source
mass. The symbols  correspond to the observed radio luminosities of superluminal
components (stars for microquasars, circles and triangles for the low luminous AGNs, i.e.,    LINERs (see below) and  Seyfert galaxies, respectively, and squares for luminous AGNs).  The same parameter space described in Sect. 3.5 was used for the AGNs, except that the black hole mass  now  spans the range
$5 M_{\odot}$ to $10^{10} M_{\odot}$.

\begin{figure}
\begin{center}
 \includegraphics[width=0.47\textwidth]{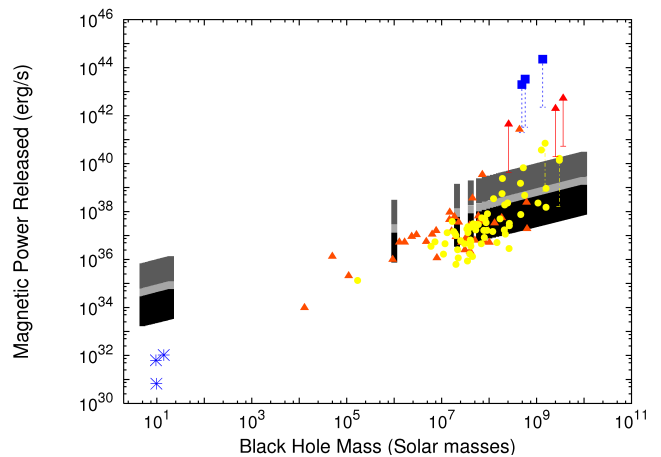}
\caption{$\dot{W}_B$ versus the BH mass $M$ for both microquasars and AGNs. The stars represent the observed radio luminosities for the same three microquasars of Fig. 2.  The circles, triangles, and squares are observed radio luminosities of jets at parsec scales from LINERS,  Seyfert galaxies, and luminous AGNs (or quasars), respectively (obtained from Kellermann et al. 1998; Nagar et al. 2005). The masses of the AGN BHs  were evaluated from the central stellar velocity dispersion given by Tremaine et al. (2002). The vertical thin bars associated to some sources stand for a reduction by a factor of 100 in the  (isotropic) observed luminosities due to  relativistic beaming. The thick bars correspond to the calculated magnetic reconnection power and encompass the parameter space that spans $5 M_{\odot} \leq M \leq 10^{10} M_{\odot}$, $0.05 \leq \alpha \leq 0.5$, $0.1 \leq \beta \leq 1$, and
$1 R_S \leq l \leq 1000 R_S$ (or $0.3 R_X \leq l \leq 333 R_X$), with $1 R_S \leq l \leq 10 R_S$ in black, $10 R_S < l \leq 30 R_S$ in light gray, and $30 R_S < l \leq 1000 R_S$ in dark gray.}
\end{center}
\end{figure}

Figure 3 indicates that the magnetic power released during violent reconnection events is able to explain the emission of relativistic radio blobs from both microquasars and the so-called low-luminous AGNs (LLAGNs), which include  Seyfert galaxies and LINERs (these latter have emission-line luminosities that are typically a factor of $10^2$ orders of magnitude smaller than those of luminous AGNs; e.g., Ho et al. 1997). This establishes a correlation between the magnetic reconnection power of stellar mass and supermassive black holes according to Eq. (12)  spanning over $10^9$ orders of magnitude in  mass of the sources.

We notice that a simple power-law dependence of $\dot{W_B}$ with the black hole mass $M$ may be derived from Eq. (12) if we consider that the height of the reconnection zone in the corona, $l$, is parameterized by $R_X$, which in turn is $R_X = 3 R_{S} =  6 GM/c^2$. This implies that according to the present coronal-disc model,  $\dot{W_B} \propto \alpha^{-19/16} \beta^{-9/16} M^{1/2}$, and this explains the approximately straight inclination of the thick gray scale curve of Fig. 3.

%The emission-line luminosities of LLAGNs
%(LH ¡Ü 1040 erg s.1 by definition; Ho et al., 1997a) are
%a factor ¡« 102 times weaker than typical SDSS AGNs.
%If LLAGNs are truly (weak) AGNs, then extending our
%studies to LLAGNs is important as they greatly outnumber
%powerful AGNs. LLAGNs are best studied in close
%(<30 Mpc) nuclei, as a result of sensitivity limitations
%and the need to attain adequate linear resolution to separate
%any weak accretion related emission from that of
%the bright host galaxy.

% We notice that the plot correlates quite well both microquasars and
% LLAGNs (low louminous AGNs). This is in agreement with the results
% of the empirically built fundamental plane  that correlates both
% liners and micorquasars spectra (Merloni et al 2003) !!

This result is consistent with the recently found empirical correlation between the radio and the (hard) X-ray luminosity, mentioned in Sect. 1, for sources spanning  from magnetically active stars to some BHXRBs and radio quiet AGNs)(Laor \& Behar 2008; see also Soker \& Vrtilek 2009). As remarked before, this correlation  clearly suggests that the radio emission in some BHXRBs and low-luminous  AGNs comes mainly from magnetic activity in the coronae above the accretion disk and is therefore nearly independent of the intrinsic
physics of the central source and the accretion disk, just like in the model above (see also de Gouveia Dal Pino 2006; de Gouveia Dal Pino et al. 2009).

%This may be do to the fact that all the activity is actually
%correlated with coronal activity around the central source and the
%accretion disk, and therefore is nearly independent of the intrinsic
%physics of the central source and the accretion disk. This is fully
%compatible with our scenario and also explains the correlation we
%have found. }

The correlation found in Fig. 3 does not hold for radio-loud AGNs, possibly because their surroundings are much denser and then "mask" the emission due to   coronal magnetic activity. In this case, particle re-acceleration behind shocks will probably prevail further out in the jet-launching region and will be the main responsible for the radio emission.

\section{In the context of YSOs}
Young stellar objects differ in many aspects from microquasars. For instance they
produce thermal rather than relativistic jets and exhibit emission
lines from which their physical properties (such as density and
temperature) are inferred. But they may also exhibit an intense
magnetic activity that results in a strong and variable x-ray
emission (e.g. Bouvier et al. 2007a, b). Observed flares in x-rays are often attributed to magnetic
activity in the stellar corona (Feigelson \& Montmerle 1999). However,
some COUP (Chandra Orion Ultra-deep Project) sources have revealed
strong flares that were related to peculiar gigantic magnetic loops
linking the magnetosphere of the central star with the inner region
of the accretion disk. It has been argued that this x-ray emission
could be due to magnetic reconnection in these gigantic loops (Favata
et al., 2005). In this section we examine this in more detail,
investigating the role of magnetic reconnection events in the inner
disc/corona of YSOs to explain the X-ray flares and check if magnetic
reconnection may be related to the thermal jets of these sources.

We note that previous numerical studies of star-disk interactions have shown that differential
rotation between the star and the inner disk regions where the stellar magnetosphere
is anchored may lead  to field lines opening and reconnection, which eventually
restores the initial magnetospheric configuration (e.g.,
Goodson \& Winglee 1999; Matt, Goodson, Winglee, et al. 2002; Uzdensky, K\"onigl
\& Litwin 2002; Romanova, Ustyugova, Koldoba, et al. 2004; von Rekowski \& Brandenburg
2004; Zanni 2009; see also Alencar 2007 for a review). Magnetospheric reconnection cycles are expected by most numerical models to
develop in a few Keplerian periods at the inner disk, and be accompanied by time dependent accretion onto
the star and by  episodic
outflow events as reconnection takes place. Also, recently several observational results have indicated that the surface
magnetic fields of young low-mass stars may be very complex and have high-order
multipoles (Valenti \& Johns-Krull 2004; Jardine, Cameron, Donati, et al. 2006; Daou,
Johns-Krull \& Valenti 2006; Donati, Jardine, Petit, et al. 2007). Accretion may then
eventually proceed through multipole field lines if the interaction region between star
and disk is not very far away from the stellar surface for only the dipole component to
be important. This is more likely to happen in stars that present high accretion rates (see below),
because them the inner gas disk will extend closer to the star than in the low mass
accretion rate systems. Multipolar axisymmetric fields were recently modeled by von
Rekowski \& Brandenburg (2006). Their time-dependent simulations include a dynamo
generated field in the star and the disk and results in a very complex and variable field
configuration. In their simulations accretion  tends to be irregular and episodic.

To analytically estimate the amount of magnetic  power that may be released by violent reconnection in the inner disk-star region we will consider here a most simple geometrical configuration to the problem as in Fig. 1. However, while the inner disk regions of microquasars are
dominated by radiation pressure, the inner disk region of YSOs
are dominated by gas pressure. Also, instead of modeling the disk/corona
of these systems, we used mean values of the coronal density and
temperature inferred from observations (Favata et al., 2005) to
parameterize our analysis.

The stellar magnetic field is assumed to be quasi dipolar in a way
that at $R_X$ this field is given by
\footnote{We note that Favata et al. (2005), assuming equipartition between gas and magnetic field,  have estimated minimum magnetic field values for the loops of the COUP sources that vary between a few dozens to a few kG. On the other hand, the  magnetic fields that are typically  measured at the surface of (class II) YSOs, i.e.,  T Tauri stars,  are between 1 and 3 kG  (e.g., Valenti \& Johns-Krull 2004 and references therein; Symington, Harries, Kurosawa, et al. 2005).
It is possible that such  strong stellar fields are mainly associated with local
multipolar components (e.g.,  starspots) and actually some of these recent polarimetric measurements
(Valenti \& Johns-Krull 2004) indicate a weaker dipolar component
(lower than 200 G). Nonetheless, because, we are  concerned with the inner disk-star interacting zone near which the amplitude of the stellar magnetic field must be maximized by the presence of small scale multipoles and compressed field lines, we  adopt stellar field magnitudes in the kG range.}
\begin{equation}
B_X = \mu B_* \left( \frac{R_*}{R_X} \right) ^3 ,
\end{equation}
where $B_*$ is the magnetic field on the surface of the star, $R_*$ is
the stellar radius and $\mu$, which is expected to assume values slightly
higher than 1, is a parameter that accounts for small deformations of the
dipole geometry due to the disk ram pressure (e.g. Shu et al 2004).

From the equilibrium between the stellar magnetic pressure and the disk
ram pressure in the inner disk region we define the radius at which the
disk is truncated (e.g., Bouvier et al., 2007b). With these assumptions, the
maximum possible accretion rate is reached when the disk touches the stellar
surface and can be estimated by
\begin{equation}
\dot{M}_{max} = 1.12\times 10^{22} \left( \mu B_{*}\right)_{5000}^2 R_2^{5/2} M_1^{-1/2} \hspace{1cm} g/s ,
\end{equation}
where $\left(\mu B_* \right) = 5000 \left(\mu B_* \right)_{5000}$ G, $R_* = 2
R_{\odot} R_2$, and $M_* = 1M_{\odot} M_1$ is the stellar mass. These high
accretion rates (i.e., around $10^{-4}$ $M_{\odot}/year$) are much higher
than the average values estimated for more evolved (class II, classical T Tauri) YSOs ($\simeq 10^{-8} - 10^{-7}$ $M_{\odot}/year$), but are on the order of the maximum rates  expected
 for younger class I objects, which may undertake  FU Ori type outbursts, for example (e.g., Hartmann 2001; Schulz 2005).

Associated to this maximum accretion rate there is a maximum magnetic power
released by magnetic reconnection events
\begin{eqnarray}
\dot{W}_{B,max}=&& 1.4 \times 10^{31} \left( \mu B_* \right)_{5000}^{-16/7} M_1^{11/7} \dot{M}_{max}^{22/7} R_{2}^{-48/7}\nonumber \\
&& \times~~n_{11}^{-3/2} T_8^{-1/2} \hspace{1cm} erg/s ,
\end{eqnarray}
 where $n_c = 10^{11} n_{11}$ $cm^{-3}$ is the coronal density, and $T_c = 10^8
T_8$ K is the coronal temperature (both estimated from observations; Favata et al. 2005).

Figure 4 shows a comparison between the predicted maximum magnetic power
released under these circumstances in violent reconnection events and the
observed x-ray luminosities (represented by star symbols) for a sample of COUP
sources (Favata et al., 2005). The different points associated with each source correspond
to different values of $\left( \mu B_* \right)$ (which imply different
values of $\dot{M}_{max}$). For most sources, the magnetic power is on the
order of (or greater than) the observed luminosities if the maximum
accretion rate is $(0.5-1)\times 10^{-4}$ $M_{\odot}/yr$ (for 10 sources) or
$(2.5 - 5) \times 10^{-4}$ (for 5 sources).

\begin{figure}
\includegraphics[width=0.5\textwidth]{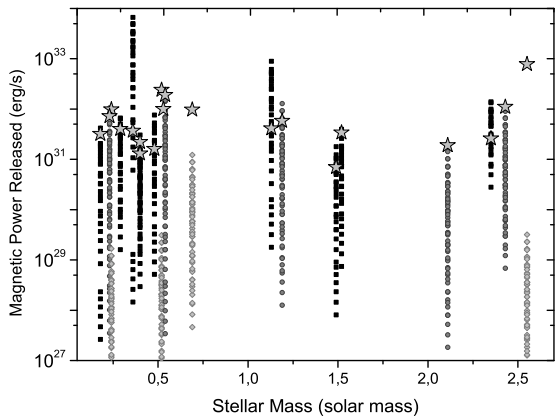}
\caption{Magnetic power released by violent magnetic reconnection events for the COUP
sources. Each vertical bar
corresponds to the calculated magnetic reconnection power for a given source mass, for magnetic fields
1 kG $\leq \mu B_{\star} \leq (\mu B_{\star})_{max}$ and accretion rates  $0.5 \dot M_{max} \leq \dot M \leq \dot M_{max}$. The lowest magnetic power value of each bar corresponds to $\mu B_{\star}= 1$ kG and  $ \dot M = 0.5 \dot M_{max}$ (eq. 16), while the highest magnetic power corresponds
to $\mu B_{\star}= (\mu B_{\star})_{max}$ (eq. 16) with
 $\dot M_{max} = 1 \times 10^{-4} M_{\odot}$ yr$^{-1}$ (black bars), and $\dot M_{max} = 5 \times 10^{-4} M_{\odot}$ yr$^{-1}$ (light-gray bars). The stars correspond to the observed X-ray power of the COUP sources
 (see text for more details).}
\label{fig:3}
\end{figure}

These accretion rates of $\sim 10^{-4}$ $M_{\odot}/yr$ about $100-1000$
times higher than the mean typical rates expected for evolved YSOs would be required only for
very short time intervals during violent magnetic reconnection events.
This implies that the detection of other spectral signatures of these
higher accretion rates would be probably very difficult.

Finally, we argue that the energy released by violent magnetic reconnection
processes could help to heat the gas at the base of the jets (Ray et al.,
2007). A rough estimate indicates that the large amount of magnetic power
released by magnetic reconnection events can be thermally conducted up to
distances $\sim 10$ AU (Pesenti et al., 2003) on a timescale $\tau _{cond}
\sim 10^8 n_{10} T_{8}^{-5/2} l_{10}^2$ s that is comparable, for instance, to the dynamical
timescale of the DGTau jet (e.g., Bacciotti et al. 2002, Cerqueira \& de Gouveia Dal Pino, 2004).

\section{Summary and conclusions}
We revisited the model proposed by de Gouveia Dal Pino
\& Lazarian (2005) for the production of relativistic ejections of
microquasar GRS 1915+195. Within this scenario, the initial
acceleration of the jet plasma to relativistic speeds is related to
violent reconnection episodes between the magnetic field lines of
the inner disk region and those that are anchored in the black
hole. This might happen whenever a large scale magnetic field (of opposite polarity to that anchored in the BH) arises from the inner accretion disk (probably amplified by dynamo action) so that the ratio between the effective disk pressure and the
magnetic pressure ($\beta$) decreases to values near 1 or below and the
accretion rate ($\dot{M}$) approaches the Eddington rate. In Sect. 3 we extended this model to other microquasars and argued that it
could be responsible for the transition from the `hard' SPLS to the
`soft' SPLS seen in other microquasars.

  In this scenario, the high accretion rate right before a radio flare results in a soft x-ray emission with a luminosity around $\sim 10^{39}$ erg/s (Sect. 3.4). The magnetic power released by violent magnetic reconnection events is mainly used to heat the coronal gas to temperatures of $\sim 10^9$ K (see Eq. 8) and heat the superficial disk gas by thermal conduction along the field lines causing  enhancement and variability of the soft x-ray flux (section 3.5). A substantial fraction of $\dot{W}_B$ is also expected to be used to accelerate particles that produce the relativistic blobs that emit radio-synchrotron-radiation with spectral index of $0.75$ if the electrons are first accelerated in the reconnection zone by  first-order Fermi like process (Sect. 3.3; de Gouveia Dal Pino \& Lazarian 2005). Further out the
relativistic fluid may be also re-accelerated behind shocks, which are expected to be
formed by the magnetic plasmons that erupt from the reconnection
zone. Behind these shocks a standard first-order Fermi acceleration must also produce  a synchrotron spectrum $S_{\nu} \propto \nu^{-0.5}$.
Both radio spectral indices are consistent with the observed
spectral range during the flares ($\alpha_{R} \simeq -0.2$ to $-1$, Dhawan et al., 2000; Hannikainen et al., 2001).
The hard x-ray emission is produced by inverse Compton scattering of the soft x-ray emission by the accelerated relativistic electrons and right after a radio flare, the hard x-ray flux should decrease, because most of these electrons are accelerated away from the disk (Sect. 3.4).

In Sect. 4 we generalized this model to AGNs. We built a diagram that compares  the calculated power released during violent magnetic reconnection versus the sources black hole masses with the observed outburst radio emission from microquasars and AGNs (Fig. 3). The results indicate  that the coronal  magnetic activity  near the jet launching region can explain the observed emission of relativistic radio blobs from microquasars to low-luminous AGNs (LLAGNs), spanning over $10^9$ orders of magnitude in  mass of the sources.
This correlation does not hold for radio-loud AGNS. These  would require reconnection events with super-Eddington accretion to explain the formation and  emission of relativistic blobs as due to reconnection. This is possibly because their surroundings are much denser and then "mask" the emission due to  coronal magnetic activity at sub-Eddington rates. In this case, particle re-acceleration behind shocks will probably prevail further out in the jet launching region and will be responsible for the radio emission.

%This super-Eddington regime would manifest itself only in a short time interval
%until the partial destruction of the magnetic flux by MR would relax this extreme
%configuration. This will be investigated in a future paper.

%Finally, we notice that the mechanism investigated here for relativistic blobs production is compatible with the proposed unified scenario for astrophysical jet production based on the magneto-centrifugal scenario (e.g., Blandford \& Payne 1982).

  A similar magnetic reconnection model applied to the jet launching region of YSO  jets suggests that  the observed X-ray flares in the sample of COUP sources (Favata et al. 2005) could be explained by violent magnetic reconnection events in the magnetosphere-inner disk coronal region only if high episodic accretion rates are achieved with $dM/dt  \simeq 100-1000 <dM/dt>$ and  the inner disk radius approaches the stellar radius during these events. These high accretion rates, however, would last only very briefly, i.e.,  not for times much longer than the typical reconnection time (Eq. 13).

  We also note that the predicted range of values for the  magnetic reconnection power of YSOs naturally lies in the left inferior part of the magnetic reconnection power versus source mass diagram of Fig. 3. This  is consistent  with the interpretation that the emission processes investigated above in all these classes of sources would  be  mainly associated with magnetic activity in the coronae and therefore would be nearly independent of the intrinsic physics of the central source and the accretion disk.

Finally, we remark that the simple analytical study carried out here has allowed us to derive  estimates of the maximum amount of energy that can be released from violent magnetic reconnection events in the inner region of several accretion disk-coronal-jet systems, as well as its potential effects on these systems, but this is only a first step, and these results must be tested with more realistic multi-dimensional numerical models which are able to capture the intrinsic non-linearity and non-steadiness character of the problem.

\begin{acknowledgements}
This work was partially supported by grants of the Brazilian Agencies FAPESP (2006/50654-3) and CNPq and by a grant of the MPIA. The authors acknowledge the useful comments of an anonymous referee.
\end{acknowledgements}
%

%\bsp

\label{lastpage}

\end{document}